\documentclass[12pt]{article}
\usepackage{subfloat}
\pdfoutput = 1
\usepackage{subcaption}
\usepackage{hyperref}
\usepackage{graphicx} 
\usepackage{amsmath}
\usepackage{amsfonts}   
\usepackage{amssymb}
\usepackage{multirow}
\usepackage{hhline}
\begin{document}
\date{Today}
\title{{\bf{\Large Black hole thermodynamics in asymptotically safe gravity }}}

\author{
{\bf {\normalsize Rituparna Mandal}$^{a}$
\thanks{drimit.ritu@gmail.com}},\,
{\bf {\normalsize Sunandan Gangopadhyay}$^{a}$\thanks{sunandan.gangopadhyay@gmail.com, sunandan.gangopadhyay@bose.res.in}}\\
$^{a}$ {\normalsize Department of Theoretical Sciences,}\\{\normalsize S.N. Bose National Centre for Basic Sciences,}\\{\normalsize JD Block, 
Sector III, Salt Lake, Kolkata 700106, India}\\[0.1cm]
}
\date{}

\maketitle

\begin{abstract}
We have investigated the black hole thermodynamics and the phase transition for renormalized group improved asymptotically safe Schwarzschild black hole. This geometry takes into account the quantum gravitational correction in the running gravitational constant identifying $G(r) \equiv G(k=k(r))$. We studied various thermodynamic quantity like the Hawking temperature, specific heat and entropy for the general parameter $\gamma$ for quantum corrected Schwarzschild metric. We have noticed that the coefficient of the leading quantum correction, that is, the logarithmic correction gets affected by the presence of $\gamma$. We further study the local temperature, thermal stability of the black hole and the free energy considering a cavity enclosing the black hole. According to the local specific heat, there exists three black hole states, among them the large and tiny black hole are thermally stable states. We further investigate the on-shell free energy and find that no Hawking-Page phase transition occurs here unlike the ordinary Schwarzschild black hole. The black hole state always prevails for all temperatures. Also, we have found two critical points, $T_{c1}$ and $T_{c2}$, corresponding to the phase transition from one black hole state to another. 
\end{abstract}

\vskip 1cm

\section{Introduction}
Black hole thermodynamics plays an important role to provide fundamental hints about the quantum aspect of gravity. Hawking and Bekenstein showed that black hole radiates with a temperature. They also found that these objects have entropy given by $S=\frac{A_{class}}{4 \hbar G}$ where $A_{hor}$ is the area of the event horizon\cite{Bekenstein_1972, Bekenstein_1973, Hawking_1974}. These findings indicate that black hole thermodynamics should be investigated in the context of quantum gravity since both Planck length and Newton's gravitational constant appear in the expression of entropy. In \cite{Bardeen_1973}, analogies to the zeroth and first law of black hole thermodynamics were found where temperature was associated with the strength of the gravitational field at event horizon called surface gravity. It was Hawking who had taken quantum effects of matter fields in curved spacetime to show that black holes radiate at a temperature and have an entropy \cite{Hawking_1975}.

\noindent A semi-classical approach is usually taken where the quantum effects are included through the matter fields and gravity is treated in a classical way. This approach breaks down at a Planck scale where quantum gravity effects become very important. To capture quantum gravity effects in thermodynamics, there are phenomenological theories, namely, the generalized uncertainty relation (GUP) \cite{Adler_2001, Dutta_2014, Dutta_2015}, modified dispersion relation with rainbow gravity \cite{Camelia_2002, Smolin_2004, Gim_2014}, tunnelling method \cite{Parikh_2000, SS_2003}. There is an alternative promising approach to a consistent quantum theory of gravitational field called asymptotic safety (AS) first proposed by Wienberg \cite{Wienberg_1979}. AS requires the existence of a nontrivial fixed point which controls the behaviour of the coupling constants. In this approach, physical quantities are safe from divergences in the ultraviolet regime, without being perturbatively renormalizable. 

\noindent The primary tool for investigating the theory is functional renormalization group equation (FRGE) \cite{Reuter_1998, Niedermaier_2006,Lauscher_2007,Codello_2009}. This idea corresponds to successively integrating out the momentum in a path integral formalism. This method permits a continuous interpolation through the renormalization group trajectories between the microscopic and macroscopic degree of freedom controlled by the effective action $\Gamma_{k}$. These trajectories satisfy a functional renormalization group equation. This results in the running of the Newton's gravitational constant.  

\noindent By setting the ``running" Newton's constant in place of Newton's constant, the renormalization group improved Schwarzschild metric has been proposed identifying $G(k) \equiv G(r)$ \cite{Reuter_2000}. Here the cut off scale $k$ is identified with a position dependent quantity, namely, the proper distance $d(r)$. The parameter $\gamma$ helps to identify the cut-off scale properly. The parameter $\gamma$ has the value $\gamma=\frac{9}{2}$ if $k=\frac{\xi}{d(r)}$ and $\gamma=0$ corresponds to $k=\frac{\xi}{r}$ \cite{Reuter_2000}.    

\noindent In this work, we will investigate the thermodynamics and phase transition for RG improved Schwarzschild metric. First, we explore the modification of the thermodynamic quantities, namely, Hawking temperature, heat capacity and entropy due to the presence of the general parameter $\gamma$. We will compare the differences in the thermodynamic quantities qualitatively and quantitatively between $\gamma=0$ and $\gamma=\frac{9}{2}$. Then, we shift attention to investigate phase transition and thermodynamic stability considering the black hole inside a finite spherical concentric cavity whose radius is larger than the horizon radius of the black hole. Here we calculate the on-shell free energy and observe that the black hole state always prevails for all temperatures incorporating quantum corrections.

\noindent The paper is organized as follows. In section 2, we describe the set up for quantum improved Schwarzschild black hole. In section 3, we study the modification of the
thermodynamic properties for RG improved Schwarzschild metric. In section 4, we investigate the phase transition and thermodynamic stability of the black hole. We conclude in section 5.

\section{Quantum improved Schwarzschild black hole}
We consider a static spherically symmetric black hole spacetime taking into account the quantum gravitational effects. For this, the key ingredient is the running Newton's constant which is obtained in the formalism of the renormalization group (RG) flow of the effective average action for gravity \cite{Reuter_1998}. The RG improved Schwarzschild metric reads \cite{Reuter_2000} 
\begin{align}
\mathrm{d}s^{2}=-f(r)\mathrm{d}t^{2}+f(r)^{-1}\mathrm{d}r^{2}+r^{2}\mathrm{d}\Omega^{2}
\label{metric}
\end{align}   
where
\begin{align}
f(r)=1-\frac{2G(r)M}{r}~.
\label{genf}
\end{align}
We now briefly review the procedure of obtaining the position dependent Newton's constant $G(r)$ from the running $G(k)$ through the identification of the infrared cut-off scale $k$ \cite{Reuter_2000}. The cut-off scale $k$ was chosen to be $k=\frac{\zeta}{d(r)}$ in terms of a position dependent function. The distance scale $d(r)$ is identified with the proper distance from the point P (Schwarzschild coordinate $(t, r, \theta, \phi)$) to the centre of the black hole along some curve at least for the spherical symmetric case. 

The approximate analytical solution for the dimensionful running Newton's constant $G(k) \equiv \frac{g(k)}{k^{2}}$  is derived to be \cite{Reuter_2000, Bonanno_2006}
\begin{align}
G(k)=\frac{G_{0}}{1+\omega G_{0} k^{2}}~.
\label{Gk}
\end{align}
Next to calculate the position dependent form of $G(k)$, one takes the form of the interpolating proper distance $d(r)$ to be \cite{Reuter_2000}
\begin{align}
d(r)=\left(\frac{r^{3}}{r+\gamma G_{0}M}\right)^{1/2}~.
\label{distance}
\end{align}
This form of $d(r)$ is chosen since for large $r$ 
\begin{align}
d(r)=r[1+\mathcal{O}(1/r)]
\label{larged}
\end{align}
and for small $r$ 
\begin{align}
d(r)=\frac{r^{3/2}}{\sqrt{\gamma G_{0} M}}+\mathcal{O}(r^{5/2})~.
\label{smalld}
\end{align}
For two curves, namely, a straight radial line from the origin to P (at fixed values of $t$, $\theta$ and $\phi$) and a spacetime curve of an observer who falls into the black hole, the proper distance for small $r$ is given by 
\begin{align}
d(r)=\frac{2}{3}\frac{1}{\sqrt{2 G_{0}M}}r^{3/2}~.
\label{properl}
\end{align} 
Comparing eq.(s) (\ref{smalld}, \ref{properl}) valid for small $r$, one can obtain the value of $\gamma=\frac{9}{2}$. We shall study the black hole thermodynamics taking the general value of $\gamma$. We shall also compare the results for two different values of $\gamma=\frac{9}{2},~0$.

\noindent Next using the eq.\eqref{distance}, the position dependent Newton's constant takes the form 
\begin{align}
G(r)&=\frac{G_{0}d(r)^{2}}{d(r)^{2}+\tilde{\omega}G_{0}} \nonumber \\ &=\frac{G_{0}r^{3}}{r^{3}+\tilde{\omega}G_{0}(r+\gamma G_{0}M)}
\label{Gdr}
\end{align}
where $\tilde{\omega}=\omega \zeta^{2}$.
Here the constants $\gamma$ and $\tilde{\omega}$ are coming from the proper cut-off identification of the infrared momentum scale $k$ in this formalism. Hence, the final form quantum corrected lapse function reads
\begin{align}
f(r)=1-\frac{2 G_{0} M r^{2}}{r^{3}+\tilde{\omega}G_{0}(r+\gamma G_{0}M)}~.
\label{fgmet}
\end{align}
 
\noindent We now first proceed to obtain the horizon of the quantum corrected Schwarzschild black hole. To determine the horizon, we write down $f(r)$ in the following form
\begin{align}
f(r)=\frac{B(x)}{B(x)+2x^2}
\label{Bx}
\end{align}
where $x \equiv \frac{r}{G_{0}M}$. The polynomial $B(x)$ is given by
\begin{align}
B(x)\equiv B_{\gamma,\Omega}(x)=x^{3}-2x^{2}+\Omega x + \gamma \Omega
\label{Bxn}
\end{align}
where
\begin{align} 
\Omega=\frac{\tilde{\omega}}{G_{0}M^{2}}~.
\label{dOmega}
\end{align}
As $\Omega$ carries the signature of the quantum corrections, the classical limit can be recovered setting $\Omega=0$. The horizon radius for the above is given by solving $f(r)=0$ which in turn implies that $B(x)=0$. For $\Omega=0$, the non-trivial classical horizon is given by $x_{0}=2$ which corresponds to the known Schwarzschild horizon $r_{0}=x_{0}G_{0}M=2G_{0}M$.

\noindent Now in the quantum case $\Omega \neq 0$, and then $B_{\gamma,\Omega}(x)$ is a cubic polynomial which has either one or three real roots. One can see that $B(-\infty)=-\infty$ and $B(0)=\gamma \Omega >0$ which infers that $B(x)$ always has at least one zero on the negative real axis. The derivative ${B}'(x)=3x^{2}-4x+\Omega$ is positive for $x<0$ which implies that $B(x)$ is monotonically increasing. Hence in the negative real axis, only one root is possible for $B(x)$. Therefore, $B(x)$ has either two real roots or no roots possible on the positive real axis. 
    
To proceed further, we define $x$ as $x=y+\frac{2}{3}$. This leads to
\begin{align}
y^3+Py+Q=0
\label{cy}
\end{align}
where $P=\Omega-\frac{4}{3}$ and $Q=\gamma \Omega +\frac{2 \Omega}{3}-\frac{16}{27}$. To solve the above cubic equation for $y$, we employ Cardano's method. We first substitute $y=u+v$ in eq.\eqref{cy} and get
\begin{align}
y^{3}-3uvy-(u^{3}+v^{3})=0~.
\label{uv}
\end{align}
Comparing eq.(s)(\ref{cy}, \ref{uv}), we obtain the following algebra identity 
\begin{align}
P=-3uv,~~~~ Q=-(u^{3}+v^{3}).
\label{PQ}
\end{align}
Substituting $u=-\frac{P}{3v}$ in $Q$, we obtain the following quadratic equation for $b^{3}$ 
\begin{align}
(v^{3})^{2}+Qv^{3}-(\frac{P}{3})^3=0.
\label{v3}
\end{align}
Now solving the above equation, the solutions of $v^{3}$ and $u^{3}$ read
\begin{align}
v^{3} &=-\frac{Q}{2}-\sqrt{(\frac{Q}{2})^{2}+(\frac{P}{3})^3}=-\frac{Q}{2}-\sqrt{\Delta} \nonumber \\
u^{3} &=-\frac{Q}{2}+\sqrt{(\frac{Q}{2})^{2}+(\frac{P}{3})^3}=-\frac{Q}{2}+\sqrt{\Delta}~~.
\label{soluv}
\end{align}
The signs before the radicals have been chosen so that the algebraic identity $Q=-(u^{3}+v^{3})$ holds.

\noindent We now define $\Delta$ as
\begin{align}
\Delta &=(\frac{Q}{2})^{2}+(\frac{P}{3})^3 =\frac{1}{108}\left(27Q^{2}+4P^{3}\right)
=\frac{1}{108}D_{\gamma, \Omega}
\label{disc}
\end{align}
where the discriminant $D_{\gamma, \Omega}$ can be written in this form 
\begin{align}
D_{\gamma, \Omega}=\frac{4}{27}\left[\left(9\Omega+\frac{27}{2}\gamma \Omega -8\right)^{2}+(3\Omega-4)^{3}\right]\,~.
\label{discri}
\end{align}
Depending on the value of the discriminant, the cubic roots are distinguished as real or complex. If $D_{\gamma, \Omega}<0$, there are two real roots, for $D_{\gamma, \Omega}>0$ no real roots exist on the positive real axis. For $D_{\gamma, \Omega}=0$, we have a double zero on the real axis. To proceed further, we observe that the discriminant can be factorized in the following form 
\begin{align}
D_{\gamma, \Omega}=\Omega\left[\Omega-\Omega_{1}(\gamma)\right]\left[\Omega-\Omega_{cr}(\gamma)\right]
\label{facdis}
\end{align}
where 
\begin{align}
\Omega_{1}(\gamma)&=\frac{1}{2}-\frac{27}{8}\gamma^{2}-\frac{9}{2}\gamma-\frac{1}{8}\left(9\gamma+2\right)^{\frac{3}{2}}\sqrt{\gamma+2} \nonumber \\ 
\Omega_{cr}(\gamma)&=\frac{1}{2}-\frac{27}{8}\gamma^{2}-\frac{9}{2}\gamma+\frac{1}{8}\left(9\gamma+2\right)^{\frac{3}{2}}\sqrt{\gamma+2}~~.
\label{omega1}
\end{align}
Now the function $\Omega_{1}(\gamma)$ is always negative for any positive value of $\gamma$. So the sign of the discriminant solely depends on the value of $\Omega_{cr}(\gamma)$. If $\Omega< \Omega_{cr}(\gamma)$, then $D_{\gamma, \Omega}<0$ so that $B_{\gamma,\Omega}(x)$ has two real roots namely $x_{+}$ and $x_{-}$ on the real positive axis. For $\Omega = \Omega_{cr}(\gamma)$, the two roots overlap into a single root at $x_{+}=x_{-}=x_{cr}$. For $\Omega > \Omega_{cr}(\gamma)$, there are no roots of $B_{\gamma,\Omega}(x)$ in the real positive axis.

\noindent Now for the horizon to exist, we therefore require to take  $\Omega< \Omega_{cr}(\gamma)$ for which the two zeros on the real positive axis take the analytical forms  
\begin{align}
x_{+}&=y_{+}+\frac{2}{3}=2\sqrt{-\left(\frac{P}{3}\right)}\cos \left ( \frac{\Theta}{3}\right )+\frac{2}{3}  \nonumber \\
x_{-}&=y_{-}+\frac{2}{3}=2\sqrt{-\left(\frac{P}{3}\right)}\cos \left ( \frac{\Theta}{3}+\frac{4\pi }{3}\right )+\frac{2}{3}
\label{xplne}
\end{align}
where $\Theta=\arccos \left (-\frac{\frac{Q}{2}}{\sqrt{-(\frac{P}{3})^{3}}} \right )$. Substituting $P$ and $Q$ in terms of $\gamma$ and $\Omega$ in $x_{\pm}$, we obtain an outer horizon $r_{+}$ and an inner horizon $r_{-}$ 
\begin{align}
r_{+}&=x_{+}G_{0}M=\frac{2}{3} \left(1+\sqrt{4-3 \Omega } \cos \left(\frac{1}{3} \cos ^{-1}\left(\frac{16-9 (3 \gamma +2) \Omega }{2 \sqrt{-(3 \Omega -4)^3}}\right)\right)\right)G_{0}M  \label{rplus}  \\
r_{-}&=x_{-}G_{0}M=\frac{2}{3} \left(1-\sqrt{4-3 \Omega } \cos \left(\frac{1}{3} \cos ^{-1}\left(\frac{16-9 (3 \gamma +2) \Omega }{2 \sqrt{-(3 \Omega -4)^3}}\right)+\frac{\pi}{3} \right)\right)G_{0}M~.
\label{rneg}
\end{align}
These solutions are for general positive $\gamma$ for which $\Omega< \Omega_{cr}(\gamma)$. We shall work with $\gamma=\frac{9}{2}$, for which the critical value is calculated from eq. \eqref{omega1} to be $\Omega_{cr}\approx 0.20$ \cite{Reuter_2000}. Now taking the limit $\Omega \rightarrow 0$ in the above solution, we obtain the classical horizon where $r_{+}=2G_{0}M$ and $r_{-}=0$. For $\gamma=0$, we have $x_{\pm}=1$ for $\Omega =\Omega_{cr}=1$. 
From the definition of $\Omega$, the critical value for the mass is calculated in terms of $\Omega_{cr}(\gamma)$, which reads 
\begin{align}
M_{cr}(\gamma)=\left[\frac{\tilde{\omega}}{\Omega_{cr}(\gamma)G_{0}}\right]^{1/2}~.
\label{critma}
\end{align}
For $\Omega=\Omega_{cr}$, we have a double zero of $f(r)$ on the positive real axis where $r_{\pm}$ coincides at $r_{cr}=x_{cr}G_{0}M$. The explicit solution for $r_{cr}$ is given in \cite{Reuter_2000}. 

\subsection{Large mass expansion of $r_{+}$} 
We are mainly interested on the outer horizon $r_{+}$ to study the black hole thermodynamics.  We shall carry out a large mass expansion of $x_{+}$ upto $\mathcal{O}(\Omega^{2})$ which in turn would give the outer horizon $r_{+}$. Starting with the classical horizon $r_{+}=2G_{0}M$, that is, $x_{+}=2$, the ansatz can be taken in the form \cite{Reuter_2000}
\begin{align}
 x_{+}=2+c_{1}\Omega+c_{2}\Omega^{2}+\mathcal{O}(\Omega^{3})~. 
\label{x+reuter} 
\end{align}
\noindent Substituting this form of ansatz for $x_{+}$ in $B_{\gamma, \Omega}(x)$ and comparing equal powers of $\Omega$ and $\Omega^{2}$, we can obtain $c_{1}$ and $c_{2}$ in the following form
\begin{align}
c_{1}=-\frac{1}{4}(2+\gamma),~~~~c_{2}=-c_{1}^{2}-\frac{c_{1}}{4}=-\frac{(2+\gamma)(1+\gamma)}{16}~.
\label{c1c2}
\end{align}
Putting $c_{1}$ and $c_{2}$ in eq.\eqref{x+reuter}, we get
\begin{align}
x_{+}=\left(2-\frac{1}{4}\left(2+\gamma\right)\Omega -\frac{1}{16}(2+\gamma)(1+\gamma)\Omega^{2}+\mathcal{O}(\Omega^{3})\right)~.
\label{xplus}
\end{align}
Substituting $\Omega$ in terms of the black hole mass $M$, the quantum corrected outer horizon $r_{+}$ in the large mass expansion reads
\begin{align}
r_{+}=x_{+}G_{0}M=2G_{0}M-\frac{\left(2+\gamma\right)\tilde{\omega}}{4M}-\frac{(2+\gamma)(1+\gamma)\tilde{\omega}^{2}}{16G_{0}M^{3}}+\mathcal{O}(\frac{1}{G_{0}^{2}M^{5}})~.
\label{xmplus}
\end{align}
Note that the quantum corrected horizon $r_{+}$ is smaller than the classical horizon.

\section{Heat capacity and entropy of the black hole}
\begin{figure}
\begin{subfigure}{.5\textwidth}
  \centering
    \includegraphics[width = 7.5 cm, height = 5.5cm]{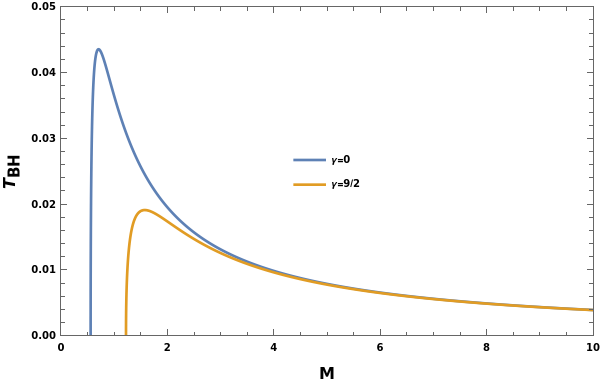}
 \caption{$T_{BH}$ vs $M$}
  \label{fig11}
\end{subfigure}%
\begin{subfigure}{.5\textwidth}
  \centering
  \includegraphics[width= 7.5 cm, height = 5.5cm]{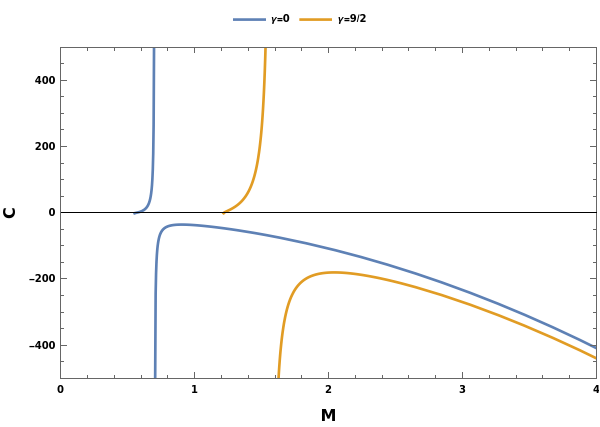}
  \caption{$C$ vs $M$ }
  \label{fig12}
\end{subfigure}%
\caption{The quantum corrected Hawking temperature and specific heat vs mass of the black hole in Planck unit.}
\label{fig1}
\end{figure}

In this section, we will explore the thermodynamic quantities such as heat capacity and entropy with general positive $\gamma$ in case of the quantum corrected Schwarzschild geometry. For this we start with the general Bekenstein-Hawking temperature of black holes which reads 
\begin{align}
T_{BH}=\frac{1}{4\pi}{f}'(r_{+})~.
\label{Tf}
\end{align}
Differentiating $f(r)$ from eq.\eqref{fgmet} at the outer horizon $r_{+}$, the quantum corrected Hawking temperature is given by \cite{Reuter_2000} 
\begin{align}
T_{BH}=\frac{1}{8\pi G_{0}M}\left[1-\frac{\Omega}{x_{+}^{2}}-\frac{2\gamma \Omega}{x_{+}^{3}}\right]~.
\label{Texact}
\end{align}
Now for $\gamma=0$ and $\Omega=\Omega_{cr}=1$, the black hole temperature is zero \cite{Reuter_2000}. We also observe that at $\Omega=\Omega_{cr}=1$, the heat capacity also vanishes for $\gamma=0$. Using $x_{+}$ from eq.\eqref{rplus} and converting $\Omega$ in terms of the mass of the black hole $M$, and using eq.\eqref{Texact}, one can obtain the Hawking temperature. The quantum corrected Hawking temperature is plotted with the mass of the black hole $M$ taking $\tilde{\omega}=0.3$ in Fig.\eqref{fig11} for $\gamma=0$ and $\gamma=\frac{9}{2}$. Note that we have set $G_{0}=1$ throughout all the plots.
From \eqref{fig11}, we can infer that for large mass black holes, the temperature almost coincides for $\gamma=0$ and $\gamma=9/2$ but when the mass of the black hole is small the temperature is relatively low for $\gamma=\frac{9}{2}$ than $\gamma=0$. Another feature that is worthy to note is that the temperature of the black hole vanishes for higher mass for $\gamma=\frac{9}{2}$ than $\gamma=0$. So the black hole critical mass or the remnant mass is large for $\gamma=\frac{9}{2}$. This can also be seen analytically by solving two equations.
The first one follows from the vanishing of the Hawking temperature of the black hole which takes zero value as it goes to critical mass and it reads
\begin{align}
x_{cr}^{3}-\Omega_{cr} x_{cr}-2\gamma \Omega_{cr}=0~.
\label{tcrite}
\end{align}
The second one is more general and follows from the vanishing of the lapse function at the event horizon which from eq.\eqref{Bxn} reads 
\begin{align}
x_{cr}^{3}-2x_{cr}^{2}+\Omega_{cr} x_{cr}+\gamma \Omega_{cr}=0~.
\label{hcrite}
\end{align}
Combining these two equations, we get
\begin{align}
2x_{cr}^{2}-2\Omega_{cr} x_{cr}-3\gamma \Omega_{cr}=0~.
\label{3crit}
\end{align}
Solving this, we get the critical radius of the black hole in terms of $\Omega_{cr}$ and $\gamma$ in the form
\begin{align}
x_{cr}=\frac{\Omega_{cr}}{2}\left[1 \pm \sqrt{1+\frac{6\gamma}{\Omega_{cr}}}\right]~.
\label{xcrit}
\end{align}
As the horizon radius $r_{cr}=x_{cr}G_{0}M$ which corresponds to double zero of $f(r)$ is always positive, $x_{cr}$ should always be positive for all values for $\gamma$. Hence one needs to take the positive sign before the square root in eq.\eqref{xcrit}. Now substituting $x_{cr}=\frac{\Omega_{cr}}{2}\left[1 + \sqrt{1+\frac{6\gamma}{\Omega_{cr}}}\right]$ in eq. \eqref{tcrite}, we get exactly the same value of $\Omega_{cr}$ in terms of $\gamma$ as in eq. \eqref{omega1}.

\noindent Substituting eq.\eqref{xplus} in the exact form of Hawking temperature (eq.\eqref{Texact}), the large mass expansion of $T_{BH}$ reads upto $\mathcal{O}(M^{-4})$
\begin{align}
T_{BH}&=\frac{1}{8\pi G_{0}M}\left[1-\frac{\Omega}{4}(1+\gamma)-\frac{\Omega^{2}}{32}(2+\gamma)(2+3\gamma)+\mathcal{O}(\Omega^{3})\right] \nonumber \\
&=\frac{1}{8\pi G_{0}M} \left[1-\frac{\Omega_{cr}}{4}(1+\gamma)\left(\frac{M_{cr}}{M}\right)^{2}-\frac{\Omega_{cr}^{2}}{32}(2+\gamma)(2+3\gamma)\left(\frac{M_{cr}}{M}\right)^{4}  +\mathcal{O}(M^{-6}) \right]
\label{Tlarge}
\end{align}
where in the second line of the equality we have used $\Omega=\Omega_{cr}\left(\frac{M_{cr}}{M}\right)^{2}$. The above expression for the temperature for large mass holds for any value of $\gamma$. It is also reassuring to note that eq.\eqref{Tlarge} reduces to the appropriate expression in the limit $\gamma \rightarrow 0$ \cite{Reuter_2000}. 

\noindent Next we define the heat capacity of the black hole which can be written identifying the internal energy of the black hole with the its total mass $M$
\begin{align}
C=\frac{\mathrm{d} M}{\mathrm{d} T_{BH}}=\left (\frac{\mathrm{d} T_{BH}}{\mathrm{d}M}\right )^{-1}=\left (\frac{\mathrm{d} T_{BH}}{\mathrm{d}\Omega}\frac{\mathrm{d} \Omega }{\mathrm{d}M}\right )^{-1}~.
\label{Hcapf}
\end{align}
Using eq.\eqref{Texact}, the heat capacity can be written in terms of $x_{+}(\Omega)$ 
\begin{align}
C=-\frac {8  \pi G_{0} M^{2}  x_{+}^{4} } { \left[4 \Omega ^2 (3 \gamma +x_{+}) x_{+}'+x_{+} \left(x_{+}^3-3 \Omega  x_{+}-6 \gamma  \Omega\right)\right]}
\label{Hexact}
\end{align}
where prime represents derivative with respect to $\Omega$. Next substituting $x_{+}$ in terms of $\gamma$ and $\Omega$ from eq.\eqref{rplus} in eq.\eqref{Hexact} where $\Omega$ is defined in eq.\eqref{dOmega}, we have plotted the specific heat capacity against the mass of the balck hole for $\gamma=0, \frac{9}{2}$. In Fig.\eqref{fig12}, the mass where the specific heat diverges is the same mass where $\frac{\mathrm{d} T_{BH}}{\mathrm{d} M}(\tilde{M}_{cr})=0$ in Fig.\eqref{fig11}. The specific heat is negative for $M>\tilde{M}_{cr}$ and goes to positive specific heat between $M_{cr}<M<\tilde{M}_{cr}$. Another point to note is that at critical mass $M_{cr}$, the specific heat also vanishes for all $\gamma$ just like the black hole temperature.

The specific heat for large mass black hole for any value of $\gamma$ reads
\begin{align}
C=-8\pi G_{0}M^{2} \left[1+\frac{3}{4}(1+\gamma)\Omega+\left(\frac{5}{32}(2+\gamma)(2+3\gamma)+\frac{9}{16}(1+\gamma)^{2}\right)+\mathcal{O}(\Omega^{3})\right]~.
\label{Heatcap}
\end{align}
In the limit $M \rightarrow \infty$, the specific heat approaches the classical value $-8 \pi G_{0} M^{2}$ which matches with the specific heat in the limit $\gamma \rightarrow 0$ \cite{Reuter_2000}.

\noindent We now proceed to calculate the entropy from the general law of thermodynamics $\left( \frac{\partial S}{\partial U}\right )_{V}=\frac{1}{T}$. In the context of black hole thermodynamics, $U$ is identified as the mass of the balck hole $M$. The differential form of the black hole entropy is then given by $\frac{\mathrm{d} S}{\mathrm{d} M}=\frac{1}{T_{BH}}$. This leads to   
\begin{align}
S=\int \frac{\mathrm{d} M}{T_{BH}}~.
\label{Sdef}
\end{align}
\begin{figure}
\begin{subfigure}{.5\textwidth}
  \centering
    \includegraphics[width = 7.5 cm, height = 5.5cm]{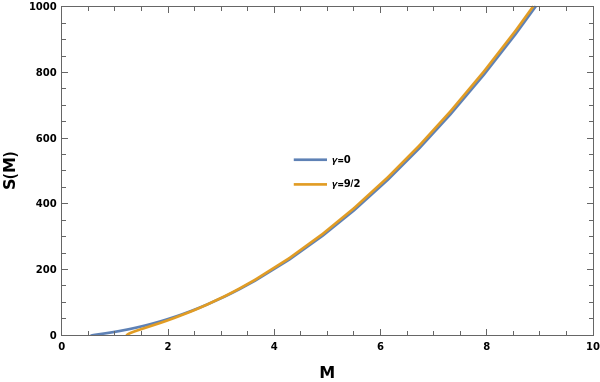}
 \caption{ $S(M)$ vs.  $M$}
  \label{fig21}
\end{subfigure}%
\begin{subfigure}{.5\textwidth}
  \centering
  \includegraphics[width= 7.5 cm, height = 5.5cm]{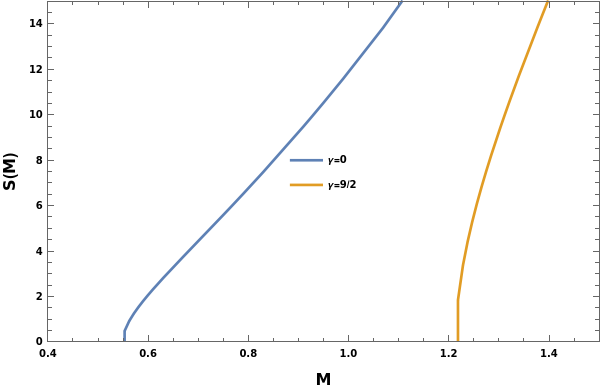}
  \caption{$S(M)$ vs. $M$ near $M_{cr}$}
  \label{fig22}
\end{subfigure}%
\caption{The entropy $S(M)$ vs. the black hole mass $M$ for $\gamma=0,\frac{9}{2}$ with $\tilde{\omega}=0.3$.}
\label{fig2}
\end{figure}
Using $x_{+}$ in terms of the black hole mass from eq.\eqref{rplus} and the black hole temperature $T_{BH}$ from eq.\eqref{Texact} in the above equation and solving it numerically, we obtain the result plotted in Fig.\eqref{fig2}.

Next we calculate the black hole entropy for the large mass black hole using the temperature from eq. \eqref{Tlarge} in the form
\begin{align}
S=\int\left[8\pi G_{0}M+2\pi\tilde{\omega}\frac{(1+\gamma)}{M}+\frac{\tilde{\omega}^{2}\pi}{4G_{0}M^{3}}\left((2+\gamma)(2+3\gamma)+2(1+\gamma)^{2}\right) +\mathcal{O}(M^{-5})\right]\mathrm{d}M~.
\label{intS}
\end{align}
Integrating the above equation, the entropy upto $\mathcal{O}(M^{-4})$ reads  
\begin{align}
S=4\pi G_{0}M^{2}+2\pi \tilde{\omega}(1+\gamma)\ln M -\frac{\tilde{\omega}^{2}\pi}{8G_{0}M^{2}}\left((2+\gamma)(2+3\gamma)+2(1+\gamma)^{2}\right)+\mathcal{O}(M^{-4})+S_{0}
\label{entropy}
\end{align}
where $S_{0}$ is the constant of integration. We can determine the integration constant $S_{0}$ by fixing  $S \rightarrow 0$ if $ M \rightarrow M_{cr}$ since $C\rightarrow0$. Then $S_{0}$ gets evaluated to
\begin{align}
S_{0}=-4\pi G_{0}M_{cr}^{2}-2\pi \tilde{\omega}(1+\gamma)\ln M_{cr} +\frac{\tilde{\omega}^{2}\pi}{8G_{0}M_{cr}^{2}}\left((2+\gamma)(2+3\gamma)+2(1+\gamma)^{2}\right)-\mathcal{O}(M_{cr}^{-4})~.
\label{Sconst}
\end{align}
In terms of the horizon area of the black hole, the expression for the entropy takes the form
\begin{align}
S=\frac{A_{class}}{4G_{0}}+\pi\tilde{\omega}(1+\gamma)\ln \left(\frac{A_{class}}{16\pi G_{0}^{2}}\right)-\frac{2\tilde{\omega}^{2}\pi G_{0}}{A_{class}}\left((2+\gamma)(2+3\gamma)+2(1+\gamma)^{2}\right)+S_{0}+\mathcal{O}(A_{class}^{-2})
\label{area}
\end{align}
where $A_{class}$ is the area defined with the classical horizon radius given by $A_{class}=4\pi (2G_{0}M)^{2}$. The first term is the classical entropy which is recovered for the heavy mass black hole. It is to be noted that the leading and subleading quantum correction terms get affected by the value of $\gamma$. The leading quantum correction term for entropy of the black hole takes higher value for $\gamma=\frac{9}{2}$ than $\gamma=0$. 

\section{Black hole mass evaporation with time}
If the ambient temperature is smaller than the the black hole temperature, the black hole radiates off its energy increasing the temperature. As a result, the mass of the black hole decreases for that case. Now from Fig.\eqref{fig1}, one can see that the qualitative nature of the Hawking temperature for large mass for $\gamma=0, \frac{9}{2}$ is similar with the classical behaviour. When the mass reaches as small as $\tilde{M}_{cr}$, the quantum corrected temperature takes the maximum value. After $\tilde{M}_{cr}$, the temperature quickly reaches zero for mass $M_{cr}$. 

\begin{figure}
\begin{subfigure}{.5\textwidth}
  \centering
    \includegraphics[width = 7.5 cm, height = 5.5cm]{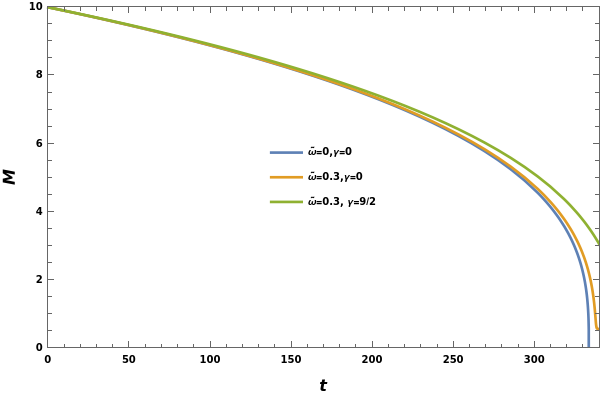}
 \caption{The mass of the black hole vs. time.}
  \label{fig31}
\end{subfigure}%
\begin{subfigure}{.5\textwidth}
  \centering
  \includegraphics[width= 7.5 cm, height = 5.5cm]{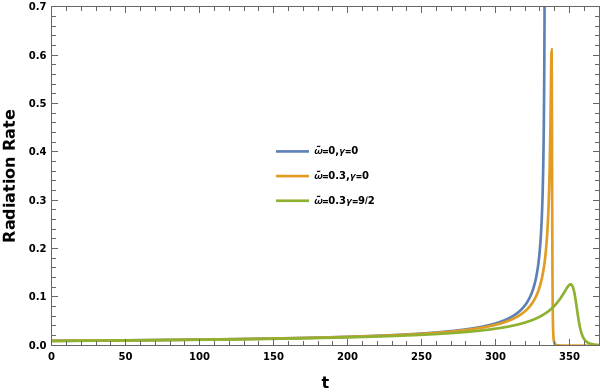}
  \caption{The radiation rate vs. time}
  \label{fig32}
\end{subfigure}%
\caption{The mass is in units of Planck mass and the time is in units of characteristic time.}
\label{fig3}
\end{figure}

Now the massive quantum corrected black hole becomes continuously evaporating increasing its temperature during the process. But the Hawking temperature unlike the classical case can never exceed beyond $T_{BH}(\tilde{M}_{cr})$. Now if one consider the loss of its energy due to photon radiation, the Stefan-Boltzmann law can be employed to calculate the mass loss as a function of time. It reads \cite{Adler_2001} 
\begin{align}
-\frac{\mathrm{d}M}{\mathrm{d}t}=\sigma \mathcal{A}(M)T_{BH}^{4}
\label{mass}
\end{align}
where $\sigma$ is Stefan-Boltzmann constant and $\mathcal{A}(M)=4\pi r_{+}^{2}=4\pi G_{0}^{2}M^{2}x_{+}^{2}$ is the area of the outer horizon. 
Substituting eq(s).(\ref{rplus}, \ref{Texact}) in the above equation, we can recast it in terms of $x=\frac{M}{M_{Pl}}$
\begin{align}
\frac{\mathrm{d}x}{\mathrm{d}t}=-\frac{x_{+}^{2}}{4 t_{ch} x^{2}}\left[1-\frac{\Omega}{x_{+}^{2}}-\frac{2\gamma \Omega}{x_{+}^{3}}\right]
\label{maeva}
\end{align}
where the characteristic time is defined as $t_{ch}=\frac{\sigma}{256 \pi^{3}G_{0}^{2}M_{Pl}^{3}}$. 
Solving the above differential equation numerically starting from some initial value $x_{i}$ (here $x_{i}=10$) upto $M \approx M_{cr}$, the mass and the radiation rate as function of time are plotted  for classical case ($\gamma=0$ and $\tilde{\omega}=0$) and for $\gamma=0, 9/2$ in Fig.\eqref{fig3}.
Here we have taken $M$ close to $M_{cr}$ for solving the mass as a function of time, as $T_{BH}$ vanishes at $M_{cr}$. It is to be noted that in Fig.\eqref{fig32}, the radiation rate decreases after $\tilde{M}_{cr}$ and reaches to zero at $M_{cr}$. As a result the time diverges if the black hole mass goes to its critical value $M_{cr}$ \cite{Reuter_2000}.

Next we proceed to calculate the approximate analytical result for the mass loss as a function of time employing the Hawking temperature and outer horizon radius for the large mass expansion. The rate at which the energy radiates is written upto $\mathcal{O}(\Omega)$ in the form
\begin{align}
\frac{\mathrm{d}x}{\mathrm{d}t}=-\frac{1}{ t_{ch} x^{2}}\left[1-\frac{(6+5\gamma)}{4}\frac{\tilde{\omega}}{x^{2}}+\mathcal{O}(\tilde{\omega}^{2})\right]~.
\label{qmass}
\end{align}
It is also to be noted that in the absence of the quantum effects, the mass evaporation rate with respect to time reads (using the fact $x_{+}=2$) \cite{Adler_2001, Dutta_2016, Dutta_2018}
\begin{align}
\frac{\mathrm{d}x}{\mathrm{d}t}=-\frac{1}{ t_{ch} x^{2}}~.
\label{maeva}
\end{align}
Solving this for initial value $x_{i}$ at $t=0$, yields the mass time relation for the classical case
\begin{align}
x=\left(x_{i}^{3}-\frac{3t}{t_{ch}}\right)^{1/3}~.
\label{classx}
\end{align}
Hence the rate at which the black hole radiates is given in terms of time as
\begin{align}
\frac{\mathrm{d}x}{\mathrm{d}t}=-\frac{1}{ t_{ch} \left(x_{i}^{3}-\frac{3t}{t_{ch}}\right)^{2/3}}~.
\label{crate}
\end{align}
The evaporation rate of mass indicates that it will stop at a time $\frac{t}{t_{ch}}=\frac{1}{3}x_{i}^{3}$ and the rate goes to infinity when the process stops for the classical black hole.

Next we proceed to carry out the above analysis for the quantum corrected black hole in the large mass expansion limit using eq.\eqref{qmass}. Solving the differential equation \eqref{qmass}, we get the solution for the mass of black hole upto $\mathcal{O}(\tilde{\omega})$ in the following form 
\begin{align}
x=\left[-\frac{3t}{t_{ch}}+ x_{i}^{3}+\frac{3(6+5\gamma)\tilde{\omega}}{4}x_{i}-\frac{3(6+5\gamma)\tilde{\omega}}{4}\left(x_{i}^{3}-\frac{3t}{t_{ch}}\right)^{1/3}+\mathcal{O}(\tilde{\omega}^{2})\right]^{1/3}
\label{quantx}
\end{align}
and the rate at which the black hole radiates is given by
\begin{align}
\frac{\mathrm{d}x}{\mathrm{d}t}&=-\frac{1}{ t_{ch} \left[-\frac{3t}{t_{ch}}+ x_{i}^{3}+\frac{3(6+5\gamma)\tilde{\omega}}{4}x_{i}-\frac{3(6+5\gamma)\tilde{\omega}}{4}\left(x_{i}^{3}-\frac{3t}{t_{ch}}\right)^{1/3}\right]^{2/3}}  \nonumber \\ & \times \left[1-\frac{(6+5\gamma)\tilde{\omega}}{4\left[-\frac{3t}{t_{ch}}+ x_{i}^{3}+\frac{3(6+5\gamma)\tilde{\omega}}{4}x_{i}-\frac{3(6+5\gamma)\tilde{\omega}}{4}\left(x_{i}^{3}-\frac{3t}{t_{ch}}\right)^{1/3}\right]^{2/3}}\right]~.
\label{qrate}
\end{align}
Note that this approximation is valid upto the mass $\tilde{M}_{cr}$ after which the quantum effect dominates.
From the rate of radiation one can conclude that the approximate time in which the black hole evaporates is given (upto $\mathcal{O}(\tilde{\omega})$) by
\begin{align}
\frac{t}{t_{ch}}=\frac{x_{i}^{3}}{3}+\frac{(6+5\gamma)\tilde{\omega}}{4}x_{i}~.
\label{solt}
\end{align}
It should be noted that the above results hold for any value of $\gamma$ lying between $0$ and $\frac{9}{2}$.

\section{Free energy and phase transition}
In this section, we move on to study phase transition for the quantum corrected asymptotically safe Schwarzschild black hole. It is known that stable thermodynamic equilibrium cannot be established in asymptotically flat spacetime. For this we enclose the black hole inside a finite concentric spherical cavity of radius $R$ which is larger than the black hole radius. The black hole within the cavity acts as a well defined thermodynamic canonical ensemble. This cavity can play a similar role as the boundary of ADS space described in \cite{Hawking_1983, Brown_1994}. To proceed further, we need to calculate the local black hole temperature in the cavity in order to obtain the local thermodynamic energy and heat capacity. The local thermodynamic analysis in terms of Euclidean Einstein action is proposed by York \cite{York_1986}.

\begin{figure}
\begin{subfigure}{.5\textwidth}
  \centering
    \includegraphics[width = 7.5 cm, height = 5.5cm]{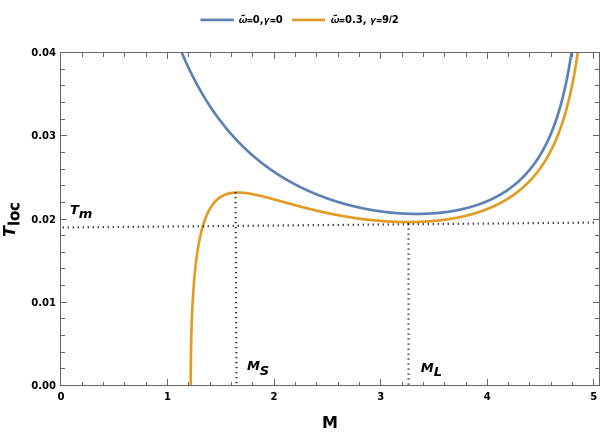}
 \caption{The local temperature vs the mass of \\ the black hole.}
  \label{fig41}
\end{subfigure}%
\begin{subfigure}{.5\textwidth}
  \centering
  \includegraphics[width= 7.5 cm, height = 5.5cm]{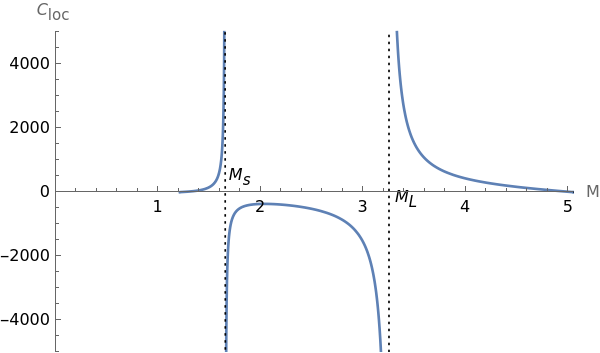}
 \caption{The local specific heat vs. the mass of the black hole ($\gamma=\frac{9}{2}$)}
  \label{fig42}
\end{subfigure}%
\caption{$T_{loc}$ and $C_{loc}$ with mass of the black hole for $\tilde{\omega}=0.3$ and $R=10$.}
\label{fig4}
\end{figure}

The local temperature $T_{loc}$ is defined on the surface of the cavity. The local temperature as seen by a local observer at $R$ is written as \cite{York_1986, Tolman_1930, Stephens_2001}
\begin{align}
T_{loc}&=\frac{T_{BH}}{\sqrt{-g_{00}}} \nonumber \\
&=\frac{\left(1-\frac{\Omega}{x_{+}^{2}}-\frac{2\gamma \Omega}{x_{+}^{3}}\right)}{8\pi G_{0}M\sqrt{1-\frac{2G_{0}MR^{2}}{R^{3}+\tilde{\omega}G_{0}(R+\gamma G_{0}M)}}}~.
\label{tloc}
\end{align} 
In Fig.\eqref{fig41}, we have plotted the local temperature vs the mass of the black hole taking $\gamma=\frac{9}{2}$, $\tilde{\omega}=0.3$ and for ordinary Schwarzschild black hole with $R=10$. As $R\rightarrow \infty$, the local temperature approaches $T_{BH}$. Also one can observe from Fig.\eqref{fig41} the existence of local minimum at $M=M_{L}$ and a local maximum at $M=M_{S}$. We now call this maximum temperature $T_{M}$ at $M_{S}$ and minimum temperature $T_{m}$ at $M_{L}$. We can see from Fig.\eqref{fig41} that for the ordinary Schwarzschild black hole (where $\gamma=0$ and $\tilde{\omega}=0$) inside a cavity of radius $R$, there exists a minimum temperature in the plot. Above this temperature, there are two values for $M$ for a given value of temperature. The black hole solution does not exist below this minimum temperature, and the radiation state (thermal AdS) prevails there. This nature exactly matches with the SAdS black hole described in \cite{Hawking_1983, Natsuume}.

The black hole geometry we are considering here exists for all possible temperatures.
From the plot we can observe that one black hole state is possible below the minimum temperature $T_{m}$, we shall call it the ``tiny black hole". Between $T_{m}<T<T_{M}$, there are three values of $M$ which in turn implies that there are three horizon radii $r_{+}$. For temperature $T>T_{M}$, only one black hole state exists, we call it the ``large black hole". Interestingly, no radiation state exists unlike the ordinary Schwarschild black hole case inside a cavity. We will again encounter these three black hole states once we calculate the specific heat capacity. Now it is well known that in Hawking-Page phase transition which occurs in SAdS black hole, if we increase the temperature, there is a phase transition from the thermal AdS spacetime (radiation state) to the SAdS black hole after a certain minimum temperature \cite{Hawking_1983, Natsuume}. In our case, if we increase the temperature, there is a phase transition from one black hole state to another. So for all temperatures, the black hole state prevails instead of the radiation state. It means there is no Hawking-Page phase transition here. We will again confirm this observation from the free energy calculation.

\noindent Next, using the first law of thermodynamics $\mathrm{d}E=T\mathrm{d}S$, one can write down the local thermodynamic energy as
\begin{align}
E_{loc}&=\int_{M_{cr}}^{M}T_{loc}\mathrm{d}S \nonumber\\ & = \int_{M_{cr}}^{M}\frac{T_{loc}}{T_{BH}}\mathrm{d}M \nonumber\\ & = \int_{M_{cr}}^{M}\frac{\sqrt{R^{3}+\tilde{\omega}G_{0}(R+\gamma G_{0}M)}}{\sqrt{R^{3}-2G_{0}M R^{2}+\tilde{\omega}G_{0}(R+\gamma G_{0}M)}}\mathrm{d}M
\label{eloc}
\end{align}
where we have used eq.(s)(\ref{Sdef}, \ref{tloc}).
Integrating the above expression, we obtain the form of $E_{loc}$ to be
\begin{eqnarray}
E_{loc}=&\frac{\sqrt{R^{3}+\tilde{\omega}G_{0}(R+\gamma G_{0}M)} \sqrt{R^{3}-2G_{0}M R^{2}+\tilde{\omega}G_{0}(R+\gamma G_{0}M)}}{\left(\tilde{\omega} \gamma  G_{0}^{2} -2 G_{0} R^{2}\right)}- \frac{2 G_{0} R^{2} \left(R^{3}+ \tilde{\omega} G_{0} R\right)} {\sqrt{\tilde{\omega} \gamma  G_{0}^{2}} \left(\tilde{\omega} \gamma G_{0}^{2}-2 G_{0} R^{2}\right)^{3/2}} \nonumber \\  & \times \sinh ^{-1}\left(\frac{\sqrt{\tilde{\omega } \gamma  G_{0}^{2} -2 G_{0} R^{2}} \sqrt{R^{3}+\tilde{\omega} G_{0}(R+\gamma G_{0} M)}}{\sqrt{2 G_{0} R^{2} \left(R^{3}+\tilde{\omega} G_{0} R \right)}}\right)-E_{loc}(M_{cr})~.
\label{eloce}
\end{eqnarray}
Note that $E_{loc}$ goes to $E_{loc}=M-M_{cr}$ in the limit $R\rightarrow \infty$. This condition is expected physically since for the quantum corrected black hole, there would be a critical mass, and therefore the amount of energy that can be extracted out of the black hole would be $(M-M_{cr})$. Now for $\tilde{\omega}=0$, the local energy goes to $E_{loc}=M$ in the limit $R \rightarrow \infty$ as $M_{cr}=0$.

To investigate the thermodynamic stability of the quantum corrected black hole, we now calculate the local heat capacity $C_{loc}$. This reads
\begin{eqnarray}
C_{loc}&=& \frac{\partial E_{loc}}{\partial T_{loc}}
=\frac{\partial E_{loc}}{\partial M} \frac{\partial M}{\partial T_{loc}}
\label{cloc}
\end{eqnarray}
\noindent In Fig.\eqref{fig42}, the local specific heat is plotted as a function of $M$. It is well known that the positive heat capacity ensure that the thermodynamic canonical system is in thermal stability. But for the negative heat capacity the black hole radiates. In Fig.\eqref{fig42}, we can observe three regions separated by discontinuity. The discontinuity or divergent point corresponds to the extrema of the local temperature. For $M>M_{L}$, the specific heat is positive which ensures that the  balck hole is thermally stable within the cavity. We call it large stable black hole (LSB). Between $M_{S}<M<M_{L}$, the negative specific heat make sure that the black hole will be in unstable state. This is called the small unstable black hole (SUB). Another region is between $M_{cr}<M<M_{s}$ where the local specific heat is again positive ensuring its thermal stability. We call it tiny stable black hole(TSB). It is to be noted that the heat capacity vanishes at non-zero value of $M$ which is identical to the critical value $M_{cr}$ calculated earlier.

\noindent With the above results in hand, we are now in a position to study the phase transition of the quantum corrected Schwarzschild black hole. For that we calculate the on-shell free energy of the black hole \cite{York_1986,Cai_2002} 
\begin{equation}
F_{on}=E_{loc}-T_{loc}S
\label{free}
\end{equation}
where $E_{loc}$, $T_{loc}$ are given in eq.(s)(\ref{tloc}, \ref{eloce})respectively. To calculate the free energy, We have used the numerical solution of the entropy $S$ using the eq.\eqref{Sdef}. It is plotted in the Fig. \eqref{fig2}. 

\begin{figure}
\centering
\includegraphics[scale=.7]{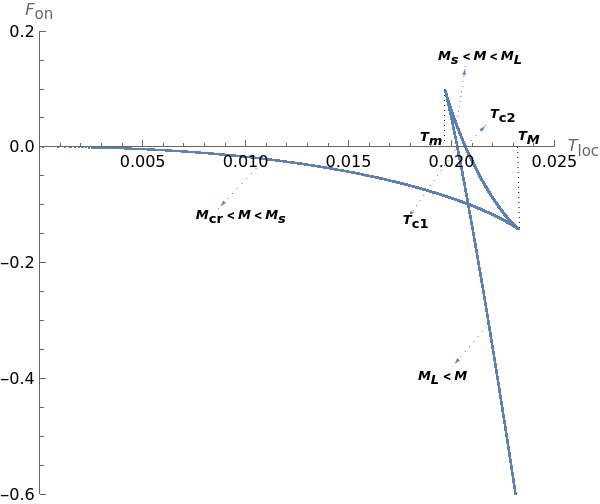}
\caption{On-shell free energy vs. the local temperature for $\gamma=\frac{9}{2}$, $\tilde{\omega}=0.3$ and $R=10$. }
\label{fig5}
\end{figure}

\noindent In Fig.\ref{fig5}, we have shown  $F_{on}$ curves with $T_{loc}$ for $\gamma=\frac{9}{2}$ using $\tilde{\omega}=0.3$ and $R=10$. Before proceeding further, we need to define the radiation state or hot curved space (HCS). The on-shell free energy vanishes for radiation state ($F_{on}=0$) for any arbitrary $T_{loc}$. This occurs as $E_{loc}=0$, $S=0$ for radiation state. In Fig. \eqref{fig5}, the $T_{loc}$ axis represents the on-shell free energy for radiation state as $F_{on}=0$ on this axis. From the plot we observe that the on-shell free energy of the radiation state is always larger than the on-shell free energy of the black hole state. Hence, the radiation state is never realizable.

\noindent Next we move on to discuss the phase transition between the three states of black hole, namely, LSB, SUB and TSB by studying their free energy. From Fig.\eqref{fig5}, one can observe that there are two critical points, namely, $T_{c1}$ and $T_{c2}$ corresponding to the phase transition from one black hole state to another.

\noindent When the temperature of the black hole ($T_{loc}$) is less than $T_{m}$, that is, $ 0 \leq T_{loc} < T_{m}$, the on-shell free energy of the tiny stable black hole $F_{on}^{TSB}$ is negative. In this region only one black hole state, that is, TSB exists. Between $T_{m}\leq T_{loc}<T_{c1}$, we have $F_{on}^{SUB}<F_{on}^{LSB}<F_{on}^{TSB}$. It implies that the small, large black hole eventually decays into the tiny black hole state when the temperature of the black hole is less than the critical temperature $T_{c1}$. For $T_{c1}<T_{loc}$, the on-shell free energy of large black hole is less than any of the state. Hence, one can conclude that below the critical temperature $T_{c1}$ the tiny black hole state (TSB) is more probable while above $T_{c1}$ the large black state is the thermally stable state.
In the temperature region between $T_{c1} \leq T_{loc}<T_{c2}$ where the three black hole exist, the on-shell free energy satisfies $F_{on}^{SUB}<F_{on}^{TSB}<F_{on}^{LSB}$. 
Here the small black hole has positive free energy so it decays into either the TSB or LSB black hole state. At $T_{c2}$, the small black hole has zero free energy value. When $T_{c2} \leq T_{loc}<T_{M}$, all three black hole states are in stable state because all three have negative free energy. Though, the on-shell free energy of the SUB and TSB are still larger than the LSB. Therefore, they undergo tunneling and eventually decays into the LSB. Finally, for $T>T_{M}$, there is only one state of the black hole possible, namely, the LSB. 

\noindent Note that for ordinary Schwarzschild black hole \cite{Gim_2014} like SAds black hole \cite{Hawking_1983}, Hawking-Page phase transition is possible which means that the radiation state prevails below a critical temperature while the large black hole state is more probable than the radiation state above this critical temperature. Remarkably, in quantum corrected Schwarzschild black hole, there is no Hawking-Page phase transition as the black hole state always prevails rather than a radition state.


\section{Conclusions}
In this paper we investigate the thermodynamics and phase transition of the renormalized group improved asymptotically safe Schwarzchild black hole. The method of renormalization group improvement for gravitational constant has been used in order to take into account the quantum gravitational effect into the spacetime. Here we first have derived the analytical solution of the outer horizon and inner horizon $r_{\pm}$ using $f(r)=0$ for general $\gamma$ depending on the value of the discriminant. From the large mass expansion, we have observed that the event horizon is smaller than the classical event horizon.  

\noindent Next we have computed the exact result for the quantum corrected Hawking temperature, specific heat and plotted it for $\gamma=0,\frac{9}{2}$. Here we have seen that upto $\tilde{M}_{cr}$, the Hawking temperature obeys the semi classical $1/M$ laws while the quantum effects are prominent after the mass $\tilde{M}_{cr}$ where the black hole temperature takes maximum value and reaches $T_{BH}=0$ at $M=M_{cr}$ (the critical or remnant mass). The specific heat diverges at $M=\tilde{M}_{cr}$ and gets negative for $M>\tilde{M}_{cr}$ but positive in between $M_{cr}<M<\tilde{M}_{cr}$. We have also calculated the Hawking temperature and specific heat for the large mass expansion for general $\gamma$. We then numerically computed and plotted the entropy of the black hole for $\gamma=\frac{9}{2},0$ and tried to report the differences in results for the different values of $\gamma$. The area theorem is recovered along with the logarithmic correction.  
Interestingly, the leading quantum correction term, that is, the logarithmic term gets bigger in the presence of $\gamma$. We further calculated and plotted the mass output and radiation rate with numerically solving the mass loss equation. We have noticed after $\tilde{M}_{cr}$, the radiation rate decreases and vanishes at the critical mass. The infinitely distant observer sees that the black hole reached its final state after an infinite time. We have also observed a analytical picture of radiation rate and mass output upto $\tilde{\omega}$ order until the mass reaches $\tilde{M}_{cr}$. 

\noindent We then moved on to investigate the phase transition and thermodynamic stability of the black hole introducing the concept of a local observer. To do this, we first calculate the local temperature, specific heat and the on-shell free energy for the quantum corrected Schwarzschild black hole. It is observed from the specific heat that three possible states of the black hole exists, namely, the large stable black hole (LSB), small unstable black hole (SUB) and the tiny stable black hole (TSB). Here we observed from the on-shell free energy that the black hole state is always more stable than the radiation state. So it is interesting to note that in quantum corrected Schwarzschild black hole, no Hawking-Page transition takes place unlike the ordinary Schwarzschild black hole. It has been also observed that there are two critical points, namely, $T_{c1}$ and $T_{c2}$ corresponding to the phase transition from one black hole state to another.

\section*{Acknowledgments} RM would like to thank DST-INSPIRE, Govt. of India for financial support.



\begin{thebibliography}{99}
\baselineskip=0.5 cm
%
%
\bibitem{Bekenstein_1972} J. D. Bekenstein, ``Black holes and the second law", Lett. Nuovo Cim. {\bf{4}}, (1972), 737–740.
%
\bibitem{Bekenstein_1973}  J. D. Bekenstein, ``Black holes and entropy", Phys. Rev. D {\bf{7}}, 2333 (1973).
%
\bibitem{Hawking_1974} S. W. Hawking, ``Black hole explosions", Nature {\bf{248}}, (1974), 30–31.

%
%
\bibitem{Bardeen_1973} J.M. Bardeen, B. Carter and  S.W. Hawking, ``The four laws of black hole mechanics", Commun. Math. Phys. 31, 161–170 (1973).
%
\bibitem{Hawking_1975} S.W. Hawking, ``Particle creation by black holes", Commun. Math. Phys.43, 199 (1975).
%
\bibitem{Adler_2001} R.J. Adler, P. Chen, D. I. Santiago, ``The Generalized Uncertainty Principle and Black Hole Remnants" Gen. Rel. Grav. {\bf{33}} (2001) 2101.

\bibitem{Dutta_2014}  S. Gangopadhyay, A. Dutta, A. Saha, ``Generalized uncertainty principle and black hole thermodynamics" Gen. Rel. Grav. {\bf{46}} (2014) 1661.
%
\bibitem{Dutta_2015} S. Gangopadhyay, A. Dutta, M. Faizal, ``Constraints on the Generalized Uncertainty Principle from black-hole thermodynamics" Euro. Phys. Lett. {\bf{112}} (2015) 20006.

\bibitem{Camelia_2002} G. Amelino-Camelia, ``Relativity in spacetimes with short-distance structure governed by an observer-independent (Planckian) length scale", Int. J. Mod. Phys. D {\bf{11}}, 35 (2002).

\bibitem{Smolin_2004}J. Magueijo and L. Smolin, ``Gravity's Rainbow" Class. Quant. Grav. {\bf{21}}, 1725 (2004)

\bibitem{Gim_2014} Y. Gim, W. Kim, ``Thermodynamic phase transition in the rainbow Schwarzschild black hole" JCAP {\bf{1410}} (2014) 003.


\bibitem{Parikh_2000} M.K. Parikh, F. Wilczek, ``Hawking radiation as tunneling", Phys. Rev. Lett. {\bf{85}} (2000).

\bibitem{SS_2003} S. Shankaranarayanan, ``Temperature and entropy of Schwarzschild-de Sitter spacetime", Phys. Rev. D {\bf{67}} (2003) 084026
%
\bibitem{Wienberg_1979}S. Weinberg, in General Relativity: An Einstein Centenary Survey, edited by S.W. Hawking and W. Israel (Cambridge University Press, Cambridge, England, 1979).
%
%
\bibitem{Reuter_1998}M. Reuter, “Nonperturbative evolution equation for quantum gravity”, Phys. Rev. D {\bf{57}}, no.10, 971 (1998).

\bibitem{Niedermaier_2006}M. Niedermaier and M. Reuter, “The Asymptotic Safety Scenario in Quantum Gravity”, Living Rev. Rel. {\bf{9}}, 5 (2006).

\bibitem{Lauscher_2007} O. Lauscher and M. Reuter, “Quantum Einstein Gravity: Towards an Asymptotically Safe Field Theory of Gravity”, Approaches to Fundamental Physics, Springer, Berlin, Heidelberg, Lecture Notes in Physics, {\bf{721}} (2007) 265-285.

\bibitem{Codello_2009} A. Codello, R. Percacci and C. Rahmede, “Investigating the Ultraviolet Properties of Gravity with a Wilsonian Renormalization Group Equation”, Annals Phys. {\bf{324}} (2009), 414-469.

\bibitem{Reuter_2000} A. Bonanno, M. Reuter, ``Renormalization group improved black hole spacetimes", Phys. Rev. D {\bf{62}} (2000) 043008.
%
\bibitem{Bonanno_2006}A. Bonanno, M. Reuter, ``Spacetime structure of an evaporating black hole in quantum gravity", Phys. Rev. D {\bf{73}}, 083005 (2006)
%
\bibitem{Dutta_2016} S. Gangopadhyay, A. Dutta, ``Constraints on rainbow gravity functions from black-hole thermodynamics" Euro. Phys. Lett. {\bf{115}} (2016) 50005.
%
\bibitem{Dutta_2018} S. Gangopadhyay, A. Dutta, ``Black Hole Thermodynamics and Generalized Uncertainty Principle with Higher Order Terms in Momentum Uncertainty", Adv. High Energy Phys. {\bf{2018}} (2018) 7450607.
%
\bibitem{Hawking_1983} S.W. Hawking, D.N. Page, ``Thermodynamics of black holes in anti-
de sitter space", Commun. Math. Phys. {\bf {87}}, 577 (1983).
%
\bibitem{Brown_1994} J. D. Brown, J. Creighton ,and R. B. Mann, ``Temperature, energy, and heat capacity of asymptotically anti-de Sitter black holes", Phys. Rev. D {\bf{50}} (1994) 6394.
%
\bibitem{York_1986} J.W. York Jr, ``Black-hole thermodynamics and the Euclidean Einstein action", Phys. Rev. D {\bf 33} (1986) 2092 .
%
\bibitem{Tolman_1930} R.C. Tolman, ``On the Weight of Heat and Thermal Equilibrium in General Relativity", Phys. Rev. {\bf{35}} (1930) 904.
%
\bibitem{Stephens_2001} G. J. Stephens and B. L. Hu, Notes on Black Hole Phase Transitions, Int. J. Theof. Phys. {\bf{40}} (2001) 2183.
%
\bibitem{Natsuume} M. Natsuume, ``AdS/CFT Duality User Guide", arXiv:1409.3575 [hep-th].

%
\bibitem{Cai_2002} R.G. Cai, ``Gauss-Bonnet black holes in AdS spaces", Phys. Rev. D 65 (2002) 084014.
%
 

%
%

%







\end{thebibliography}
\end{document}